\documentclass{article}

\usepackage{microtype}
\usepackage{graphicx}
\usepackage{subfigure}
\usepackage{booktabs} 
\usepackage{hyperref}

\usepackage[accepted]{icml2024}

\usepackage{amsmath}
\usepackage{amssymb}
\usepackage{mathtools}
\usepackage{amsthm}
\usepackage{nicefrac}
\usepackage{mathrsfs}
\usepackage{braket}
\usepackage{wrapfig}
\usepackage[normalem]{ulem}

\usepackage[capitalize,noabbrev]{cleveref}

\theoremstyle{plain}

\theoremstyle{definition}

\theoremstyle{remark}

\expandafter\def\expandafter\normalsize\expandafter{%
    \normalsize%
    \setlength\abovedisplayskip{4pt}%
    \setlength\belowdisplayskip{8pt}%
    \setlength\abovedisplayshortskip{-8pt}%
    \setlength\belowdisplayshortskip{2pt}%
}

\usepackage[textsize=tiny]{todonotes}

\newcommand{\curl}{\nabla \times}

\newcommand{\E}{\mathcal{E}}
\renewcommand{\H}{\mathcal{H}}
\newcommand{\D}{\mathcal{D}}
\newcommand{\R}{\mathbb{R}}
\newcommand{\B}{\mathcal{B}}
\newcommand{\J}{\mathcal{J}}

\newcommand{\mbf}[1]{\mathbf{#1}}

\icmltitlerunning{Nonlinear Computation with Linear Optics via Source-Position Encoding}

\begin{document}

\twocolumn[
\icmltitle{Nonlinear Computation with Linear Optics via Source-Position Encoding}

\begin{icmlauthorlist}
\icmlauthor{Nick Richardson}{Princeton}
\icmlauthor{Cyrill B{\"o}sch}{Princeton}
\icmlauthor{Ryan P. Adams}{Princeton}
\end{icmlauthorlist}

\icmlaffiliation{Princeton}{Department of Computer Science, Princeton University}

\icmlcorrespondingauthor{}{njkrichardson@princeton.edu, cb7454@princeton.edu}
\icmlkeywords{Machine Learning, ICML}

\vskip 0.3in
]

\printAffiliationsAndNotice{} 

\begin{abstract}
Optical computing systems provide an alternate hardware model which appears to be aligned with the demands of neural network workloads. However, the challenge of implementing energy efficient nonlinearities in optics -- a key requirement for realizing neural networks -- is a conspicuous missing link. 
In this work we introduce a novel method to achieve nonlinear computation in fully linear media. 
Our method can operate at low power and requires only the ability to drive the optical system at a data-dependent spatial position. 
Leveraging this positional encoding, we formulate a fully automated, topology-optimization-based hardware design framework for extremely specialized optical neural networks, drawing on 
modern advancements in optimization and machine learning. 
We evaluate our optical designs on machine learning classification tasks: 
demonstrating significant improvements over linear methods, and competitive performance when compared to standard artificial neural networks. 
\end{abstract}

\section{Introduction}
\label{introduction}

Modern machine learning systems demand an unprecedented scale of resources to drive application performance \cite{radford2019language,brown2020language}. 
Sustaining these efforts requires infrastructure developments, or improvements in efficiency; most likely both. 
The resulting emphasis on domain-specific hardware \cite{jouppi2017datacenter}, coincident with the rampant popularity of neural networks for machine learning applications, has sparked 
a resurgence of interest in alternative hardware platforms for neural network workloads \cite{schuman2022opportunities}. 

Computing with optics is particularly attractive, since optical signals are well-aligned with parallel, high-bandwidth-requiring dataflows, like those found in many 
neural networks \cite{sui2020review}. 
The linear scattering phenomena inherent to electromagnetic energy propagation is amenable to efficient implementations of the multiply-accumulate-based kernels which dominate neural network workloads \cite{ivanov2021data}. 
All that said, a critical unsolved problem is achieving efficient \textit{nonlinearities} in optical hardware.  

Conventional materials-based optical nonlinearities typically require some combination of exotic materials, difficult fabrication procedures, or high-power light sources which negate the efficiency gained through more-or-less passive linear wave propagation \cite{zuo2019all, feldmann2021parallel, keyes1985optical}. 
An alternative is to give up on an all-optical design, and compromise with optoelectronic transduction; essentially executing linear computational kernels in the optical domain, and nonlinearities in standard electronics \cite{hughes2018training, george2019neuromorphic}. 
This strategy seems to use both media in their `ideal' operating regime, but much like analog electronic neural networks, these systems are ultimately gated by fundamental inefficiencies involved in transducing signals between the two physical media \cite{rekhi2019analog}. 

As a result, the past year has seen several works concerned with developing low-power methods for nonlinear computation using linear media \cite{xia2024nonlinear, li2024nonlinear, wanjura2024fully, yildirim2024nonlinear, momeni2023backpropagation}; largely with an emphasis on machine learning applications.
A discussion on prior art in inverse design for computational electromagnetics, and background on nonlinear optics are provided in \cref{appendix:related_work:photonic_inverse_design} and \cref{appendix:related_work:nonlinear_optics}.
We discuss recent efforts toward nonlinear computation with linear optics, and its relationship to our own work in \cref{appendix:related_work:optical_neural_networks}. 

In this work, we contribute a novel method for nonlinear computation using low power, fully linear optics: source positional encoding. 
We exploit the fact that the relationship between the measured fields and the position of the source is nonlinear, and encode the data directly in the source position. 
Then, we optimize the spatial distribution of material such that the source-position-to-measured-field approximates a desired transfer function (e.g., a classifier). 
By harnessing a custom, differentiable full-wave field solver in concert with topology optimization for design of the optical hardware, we avoid analytic simplifications of the physics beyond discretization of Maxwell's equations (a source of error which is well understood and largely controllable).

In \cref{source_positional_encoding}, we detail the fundamental physics underlying our position-based encoding method. 
\cref{proposed_implementation} describes a proposed system design (see \cref{fig:system_design}) for an optical classifier leveraging positional encoding as a nonlinearity. 
\cref{in_silico_inverse_design} describes the inverse problem associated with the hardware design; which is addressed with topology optimization and differentiable simulation. 
\cref{experiments} presents experimental results across an array of test problems, with an emphasis on ablations verifying the effectiveness of our optical nonlinearity over fully linear methods. 
Finally, \cref{limitations_and_future} offers limitations of our method and promising future directions. 

\section{Method}
\label{method}

\subsection{Source positional encoding}
\label{source_positional_encoding}
Consider Maxwell's equations\footnote{Our notation, and a brief review of the relevant electromagnetics is reviewed in \cref{appendix_em}.} in the frequency domain, denoting the electric field with $\mathcal{E}$, and the source with $\mathcal{J}$. 
The resultant wave equation over a domain $\Omega \subset \R^3$ and subject to Dirichlet boundary conditions can be written as
\begin{align}\label{eq_maxwell_continuous_cy}
\curl \mu_r^{-1}(\mbf{z}) \curl \mathcal{E}(\mbf{z},\omega) - \omega^2\epsilon_r(\mbf{z}) \mathcal{E}(\mbf{z},\omega) &= j\omega\mathcal{J}(\mbf{z},\omega)
\\ \mathbf{z} \in \Omega, \quad \mathcal{E}(\mbf{z},\omega) = 0 \quad &\forall \mathbf{z} \in \partial \Omega. 
\end{align}
The medium is modeled as isotropic and lossless, i.e. the relative permeability $\mu_r$ and relative permittivity $\epsilon_r$ are real-valued scalars.
The corresponding generalized eigensystem is: 
\begin{align} \label{eq_generalized_eigenvalue}
    \curl\mu_r^{-1}(\mbf{z})\curl\mathcal{E}(\mbf{z},\omega) = \omega^2\epsilon_r(\mbf{z}) \mathcal{E}(\mbf{z},\omega).
\end{align}
We further define the following inner product:
\begin{equation} \label{eq_hermitian_inner_product}
    \langle \mathcal{E}(\mbf{z},\omega), \mathcal{E}'(\mbf{z},\omega) \rangle_{\epsilon_r} := \int_{\Omega} \mathcal{E}^*(\mbf{z},\omega)\epsilon_r(\mbf{z}) \mathcal{E}'(\mbf{z},\omega) \; d\mbf{z}.
\end{equation}
The wave operator $\mathcal{M} := \curl \mu_r^{-1} \curl$ is self-adjoint with respect to this inner product. 
Therefore, by the spectral theorem \cite{hermite1855remarque, cauchy1829equation}, $\mathcal{M}$ has a discrete spectrum with eigenvalue-eigenfunction pairs $(w_i,\mathcal{E}_i)$.  
Suppose the eigenfunctions are chosen to be orthonormal with respect to the inner product \eqref{eq_hermitian_inner_product}, i.e. $\langle \mathcal{E}_i(\mbf{z}), \mathcal{E}_j(\mbf{z}) \rangle_{\epsilon_r} = \delta_{ij}$, such that $\{\mathcal{E}_j\}_{j = 1}^{\infty}$ forms an orthonormal basis. 

\begin{figure}[hbt!]
    \includegraphics[width=0.4\textwidth]{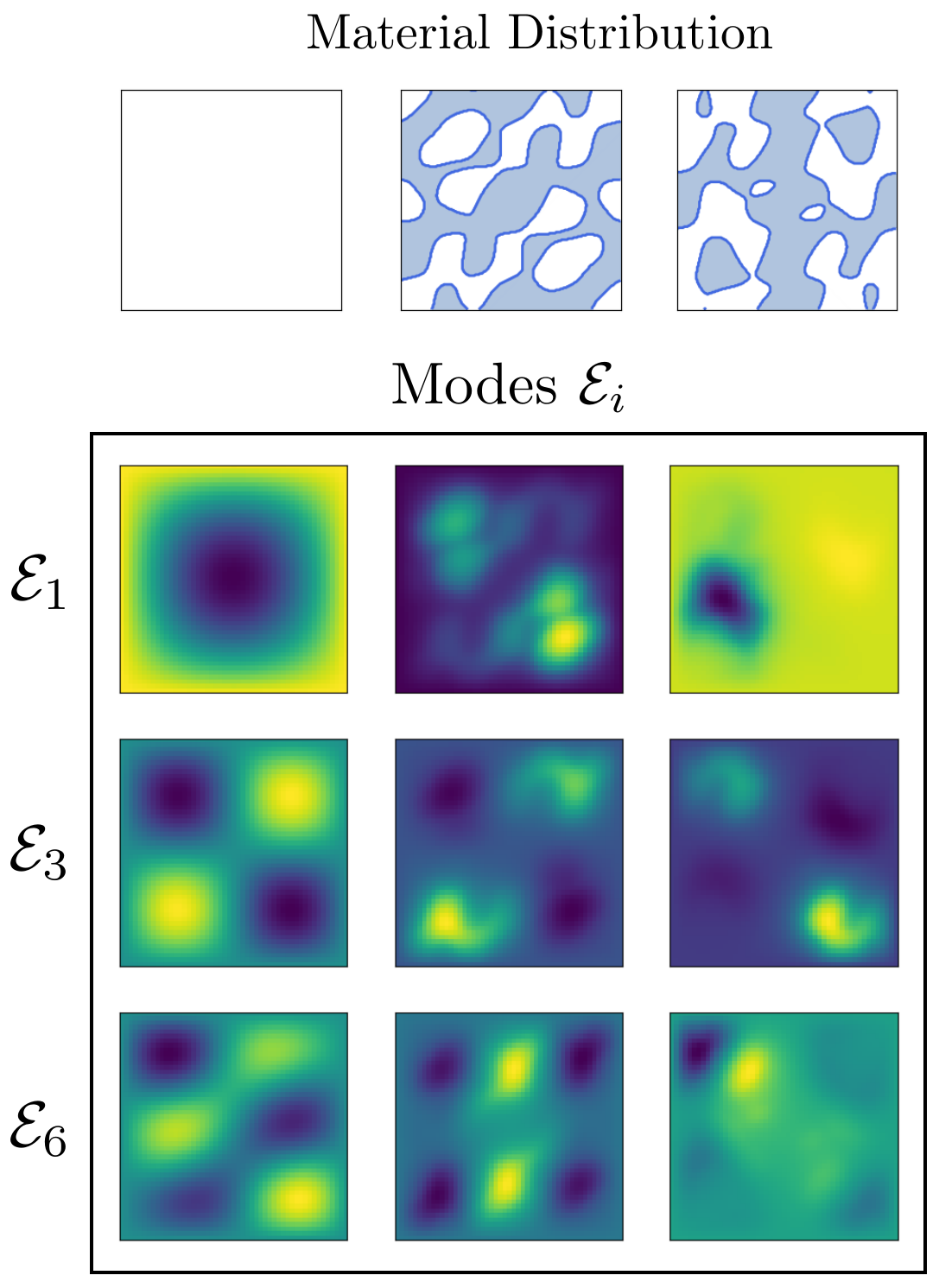}
    \vspace{-0.25cm}%
    \caption{
    Three material distributions (vacuum with $\epsilon_r = 1$, and two arbitrarily chosen distributions, with $\epsilon_r = 3$ in blue regions and $\epsilon_r = 1$ in white regions; $\mu_r = 1$ for all of them) and the 1st, 3rd and 6th corresponding 
    eigenmodes.
    Clearly, the mode profiles 
    can be quite distinct from the basic freespace modes (and one another) by varying the material distribution. 
    Complex material distributions give rise to complex modes, implying a highly nonlinear relation between source position and measured field.}
    \label{fig:modes}
\end{figure}

\begin{figure*}[hbt!]
    \includegraphics[width=\textwidth]{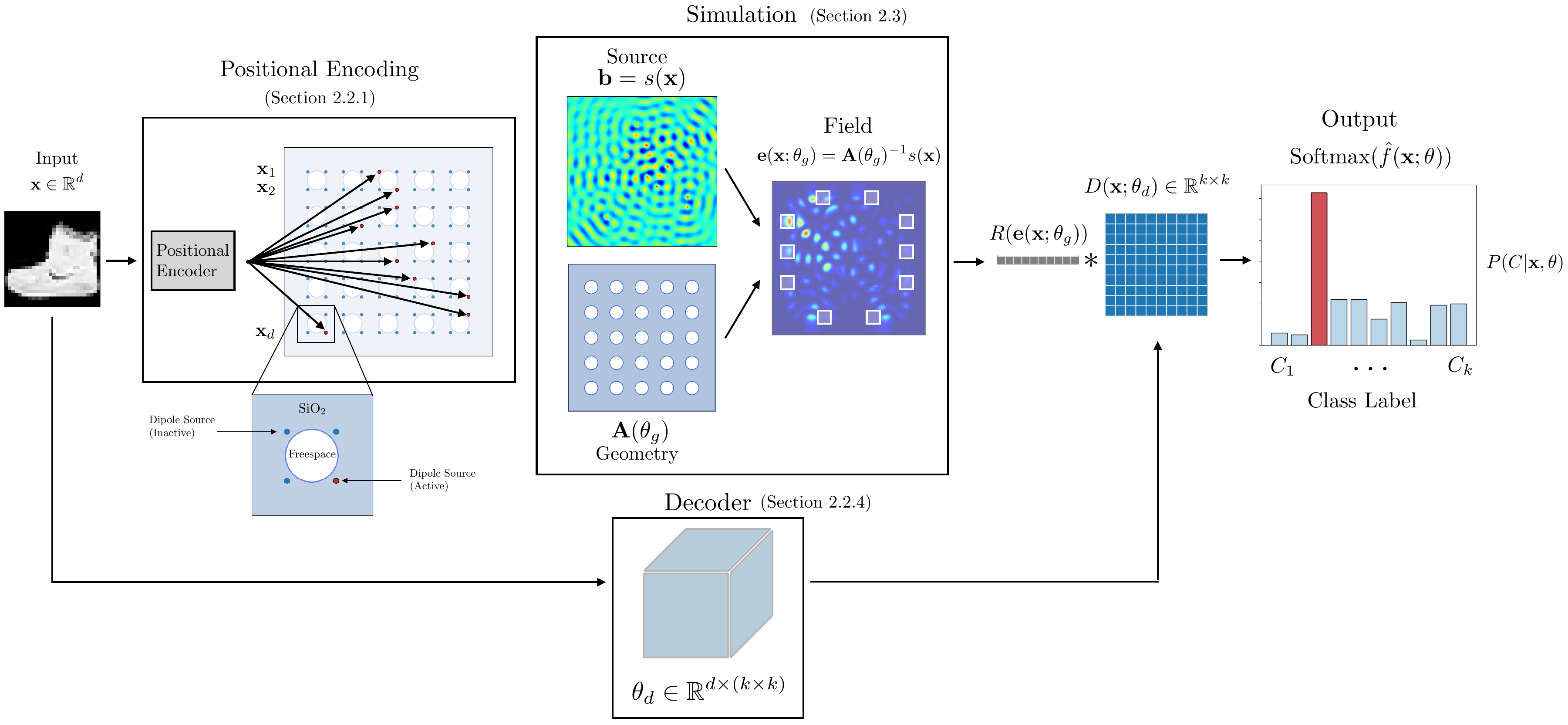}
    \vspace{-0.6cm}%
    \caption{The input data $\mathbf{x}$ is used to determine which subset of the array of sources are active (illustrated in red). This collection of active sources is assembled into an aggregate electromagnetic 
    source $\mathbf{b} := s(\mathbf{x})$. Below the aggregate source field we illustrate the geometry (hardware) $\mathbf{A}(\theta_{g})$, which does not depend on the data 
    but is optimized via a gradient method. The source and geometry are simulated in a differentiable field solver, resulting in the field shown. At a collection of fixed locations (shown as white 
    rectangular regions imposed on the field), the energy is summed and re-mapped via a data-dependent linear decoding $D(\mbf{x}; \theta_d)$. In a classification application, the result is normalized and interpreted as the readout probabilities of the system (see rightmost).}
    \label{fig:system_design}
\end{figure*}

Then it follows that any field $\mathcal{E}(\mbf{z},\omega)$ satisfying \eqref{eq_maxwell_continuous_cy} can be decomposed as, 
\begin{equation} \label{eq_eigen_decomp}
    \mathcal{E}(\mbf{z},\omega) = \sum_{i} c_i(\omega) \mathcal{E}_i(\mbf{z}). 
\end{equation} 
Inserting this decomposition into \eqref{eq_maxwell_continuous_cy}, using \eqref{eq_generalized_eigenvalue} and computing the inner product \eqref{eq_hermitian_inner_product} with the eigenfunction $\mathcal{E}_m$ results in \cite{gilbert1971excitation, ulrich2022diffuse}: 
\begin{equation}
    c_m(\omega)\omega_m^2 = \omega^2 c_m(\omega) -j\omega \braket{\E_m(\mathbf{r})|\J(\mathbf{r},\omega)}_{\epsilon}.   
\end{equation}
The coefficients can be written explicitly as
\begin{equation} \label{eq_analytic_coefficient}
    c_m(\omega) = \frac{-j\omega}{\omega_m^2 - \omega^2} \langle \mathcal{E}_m(\mbf{z}), \mathcal{J}(\mbf{z},\omega) \rangle_{\epsilon_r}.
\end{equation}
For simplicity, consider the source to be a monochromatic point source at location $\mbf{z}_s \in \Omega$ with frequency $\omega_s$ and unit amplitude, i.e. $\mathcal{J}(\mbf{z},\omega) = \delta(\mbf{z} - \mbf{z}_s)\delta(\omega-\omega_s)$. 
Then $c_i = \frac{-j\omega_s}{\omega_i^2 - \omega_s^2} \mathcal{E}^*_i(\mbf{z}_s)$. 
The electric field at a measurement location $\mbf{z}_R$ is therefore given by
\begin{equation}
    \label{eq:nonlinear_field}
    \mathcal{E}(\mbf{z}_R) = \sum_i \frac{-j\omega_s}{\omega_i^2 - \omega_s^2} \mathcal{E}^*_i(\mbf{z}_s)\mathcal{E}_i(\mbf{z}_R).
\end{equation}
This can be interpreted as the statement that the field measured at $\mbf{z}_R$ (a fixed location) is a linear combination of modes $\mathcal{E}_i$ whose amplitudes depend on (i) the deviation between source and mode frequency, and (ii) the relative spatial placement of the source and the mode amplitude profile. 

Concretely, consider an eigenfunction $\mathcal{E}_m$, and suppose it is non-degenerate (the multiplicity of $\omega_m$ is $1$), with source frequency $\omega_s$ close to its associated eigenfrequency $\omega_m$. 
Then the field is dominated by this particular mode:
\begin{equation}
    \mathcal{E}(\mbf{z}_R) \approx \frac{-j\omega_s}{\omega_m^2 - \omega_s^2} \mathcal{E}^*_m(\mbf{z}_s)\mathcal{E}_m(\mbf{z}_R).
\end{equation}
Holding $\mathbf{z}_R$ constant, and varying $\mathbf{z}_s$, the measured field follows the functional form of $\mathcal{E}_m$; which is nonlinear. 
In freespace, $\mathcal{E}_m$ is sinusoidal (\cref{fig:modes}, left column), but with general heterogeneous media, $\mathcal{E}_m$ can have a nearly arbitrarily complicated functional form (\cref{fig:modes} second and third column from the left). 
By choosing the distribution of material in the domain through optimization, we can achieve targeted mode profiles, in turn realizing a desired input to output map $\mbf{z}_s \mapsto \E(\mbf{z}_R)$.

In the general case, multiple modes whose frequencies are close to the source frequency will be excited, and the relation $\mathbf{z}_s$ to $\mathcal{E}(\mbf{z}_R)$ will be a linear combination of the modes. 

\begin{figure*}[t!]
    \includegraphics[width=0.85\textwidth]{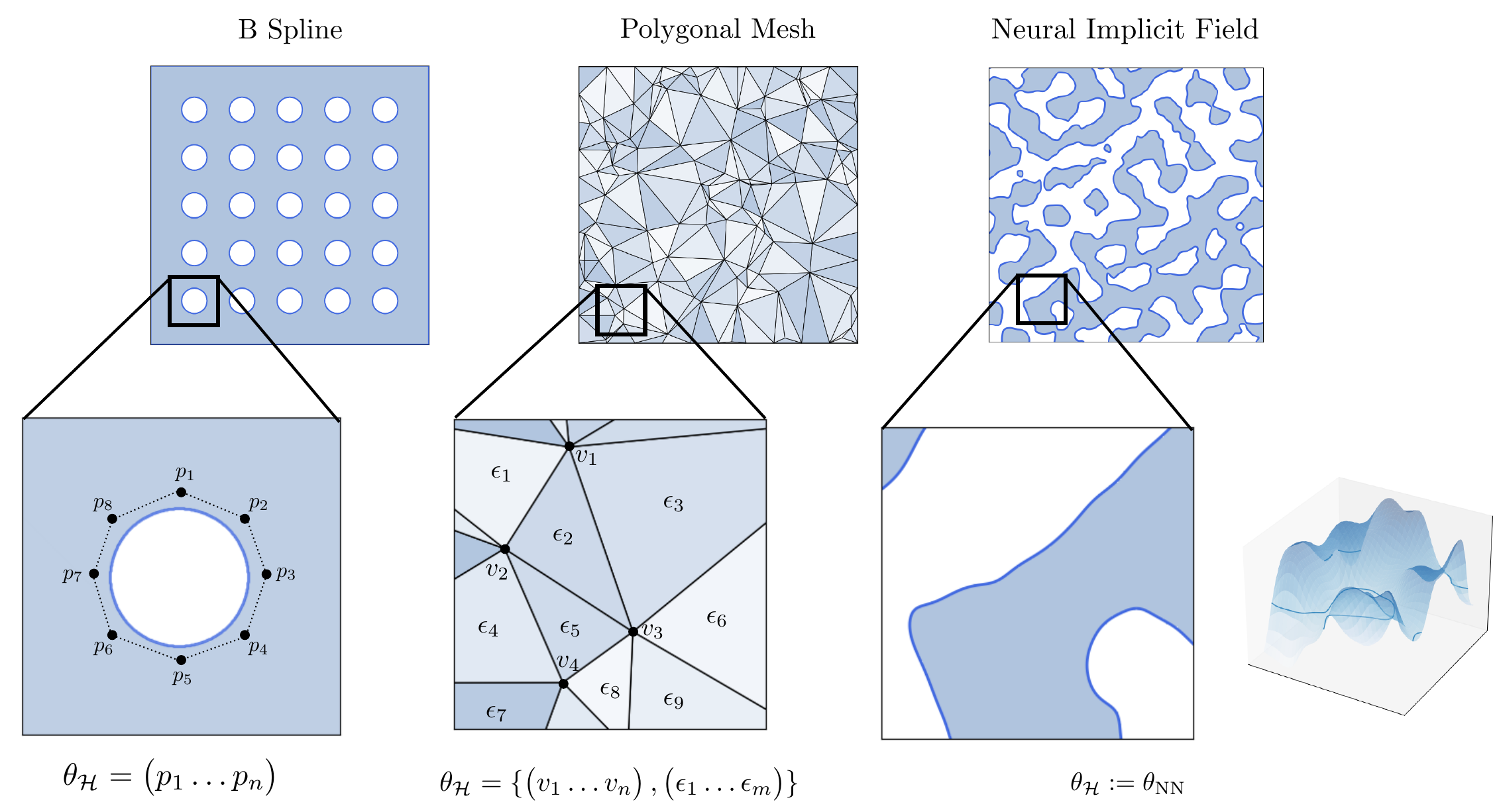}
    \vspace{-0.5cm}%
    \caption{\small Hardware parameterization instances. On the left, we show B spline patches whose interior determine freespace pores amidst a homogeneous dielectric (e.g., SiO$_2$). The parameters $\theta_{\mathcal{H}}$ modulate the shape 
    of the pores through the spline control points $p_1, \dots, p_n$. In the center, a triangular mesh is parameterized by a collection of $n$ vertices $v_1, \dots, v_n$, and permittivities $\epsilon_1, \dots, \epsilon_m$ within the interior of each triangle in the mesh. In principle, the intermediate permittivities could be quantized into several 
    discrete values to minimize the number of materials required. On the right, a neural implicit field parameterizes the distribution of material via its zero sublevel set. Details of this implicit parameterization are in (cref appendix).}
    \label{fig:hardware_parameterizations}
\end{figure*}

\subsection{Proposed Implementation}
\label{proposed_implementation}

While positional encoding is a generic method to exploit linear systems for nonlinear computation, here we consider a concrete application and describe a proposed hardware design for an optical classifier, illustrated in \cref{fig:system_design}. 

The proposed system is functionally comprised of four stages: encoding, optical dataflow, readout, and decoding. 

\subsubsection{Source Encoding}
\label{source_encoding}
Consider the source encoding system illustrated on the left in \cref{fig:system_design}, with a regular array of sources of size $d \times k$. 
While any sufficiently local source can be chosen, here we model a generic dipole emitter, i.e. a cylindrical source with common driving frequency. 
We logically partition the source array by row, so that the $i$\textsuperscript{th} row corresponds with the $i$\textsuperscript{th} entry of the input $\mbf{x} \in \mathbb{R}^d$. 
A straightforward method for positional encoding is to uniformly quantize $\mbf{x}_i$ to $k$ distinct values (i.e., $\log(k)$ bits of precision); and then power only the source in row $i$ corresponding with the quantized value. 
That is, for any input $\mbf{x}$, precisely one source is active per row of the source array. 

\subsubsection{Optical Dataflow}

Given the source $s(\mbf{x})$ derived from the positional encoding methodology above, executing the `forward pass' of the optical neural network is equivalent to simply driving the material design with the source, i.e. the computation is physically realized by the resultant wave propagation. 

We denote the material distribution obtained from optimization as $\mu^*_r$ and $\epsilon^*_r$.
Then the wave operator associated with the optimized hardware design is given by: \begin{equation}
\mathcal{A}(\mathbf{z},\omega; \mu^*_r,\epsilon^*_r) :=-\frac{j(\curl \mu_r^{*^{-1}}(\mathbf{z}) \curl - \omega^2\epsilon^*_r(\mathbf{z}))}{\omega}, 
\end{equation}
and the field is given by the solution to \eqref{eq_maxwell_continuous_cy}, i.e. \mbox{$\mathcal{E} = \mathcal{A}^{-1}s(\mathbf{x})$}.

\subsubsection{Readout}
\label{readout}

We measure the intensity of the electric field in small volumes centered at locations $\{\mathbf{z}_k\}_{k=1}^K$, for instance with avalanche photodiodes. 
This gives rise to $K$ non-negative scalar read outs. 
In our device model, we denote this operation with $R(\mathcal{E}) = R(\mathcal{A}^{-1}s(\mathbf{x})) \in \mathbb{R}_{\succeq 0}^K$. 

\subsubsection{Decoding}
\label{decoding_intra}

Before normalizing $R(\mathcal{E})$ to a probability vector, we execute a decoder, $D$, with parameters $\theta_d$ that are fit during training and fixed at inference time. 

The decoder is a strictly linear map which accounts for the fact that the association between readout location $\mbf{z}_k$ and class label is chosen arbitrarily at initialization. 
Concretely, this provides a form of invariance; if we were to permute the class labels arbitrarily, applying that same permutation to the rows of $D(\mbf{x}; \theta_d)$ would keep our predictions fixed.

While the linear restriction could be relaxed, here it is crucial to certify that the nonlinearity is a result of the source position encoding, and not the decoder. 

In summary, the aggregate system can be modeled by the following map: 
\begin{align} \label{eq_optical_network_function}
f(\mbf{x}; \omega, \mu, \epsilon, \theta) := D(\mbf{x}; {\theta_d})R(\mathcal{A}(\omega;\mu,\epsilon)^{-1}s(\mbf{x}))
\end{align}

\begin{table*}[t!]
    \begin{center}
    \scalebox{1}{
    \begin{sc}
    \begin{tabular}{llllll}
    \toprule\\
         [-1.5ex]
          \multicolumn{1}{c}{\bf Dataset}  &\multicolumn{1}{c}{\bf Quantization (bits)}  &\multicolumn{1}{c}{\bf Linear Model}  & \multicolumn{1}{c}{\bf Optical Network } & \multicolumn{1}{c}{\bf ANN (MLP) }  \\[0.7ex]
            \hline \\[-1.5ex]
          MNIST\footnote{See (cref Lecun).} & 8 & 48.6\% & 56.0 \% (\small{\textbf{+15.2\%}}) & 56.6 \%  \\
          MNIST & 24 & 75.8 \% & 83.1\% (\small{\textbf{+9.6\%}})  & 89.1\%  \\
          MNIST & 64 & 84.8\% & 86.7\% (\small{\textbf{+2.2\%}}) & 96.9\% \\
          FashionMNIST & 8 & 47.1\% & 53.1\% (\small{\textbf{+12.7\%}}) & 54.7\%  \\
          FashionMNIST & 24 & 64.1\% & 69.6\% (\small{\textbf{+8.6\%}}) & 72.4\%  \\
          FashionMNIST & 64 & 75.6\% & 79.1\% (\small{\textbf{+4.6\%}}) & 82.1\% \\
    \bottomrule
    \end{tabular}
    \end{sc}
    }
    \caption{\small Experimental results for benchmark datasets. The linear model and MLP are fit numerically using the same quantized data as in the optical neural network. In bold, we provide the relative percentage improvement (over linear) for the optical network configurations.   
    The optical networks optimize the topology in a design space of $500 \times 500$, the results cited here are from the neural field parameterization. Additional training and experiment details can be found in \cref{appendix:experiments} 
    }
    \label{table:experimental_results}
    \end{center}
    \end{table*}
    
\subsection{In Silico Inverse Design}
\label{in_silico_inverse_design}

The system model above defines a family of classifiers parameterized by $(\theta_d, \mu, \epsilon)$.
Training a classifier amounts to making a particular choice of these parameters, which we obtain by constructing a differentiable computational model for \cref{eq_optical_network_function} and using gradient based optimization.

 We discretize the wave operator $\mathcal{A}\rightarrow \mathbf{A} \in \mathbb{C}^{n^2 \times n^2}$ and the source $s(\mathbf{x})\rightarrow \mathbf{b}({\mathbf{x}}) \in \mathbb{C}^{n^2 }$ on a two-dimensional rectilinear grid of size $n \times n$ using the finite difference frequency domain (FDFD) method (see \cref{appendix_em} for details). 
 Then, we use topology optimization in tandem with a custom differentiable simulator to formulate a gradient-based co-design methodology for the material distribution and the decoder parameters. 

\subsubsection{Topology Optimization}
\label{in_silico:topopt}

We formulate the problem of material distribution via level set topology optimization \cite{van2013level}.   
The idea is to frame the choice of a particular spatial distribution of material as an optimization problem, whose objective, 
in this context, is a given function of the electromagnetic fields which depends on the material distribution through the wave operator. 

Given a rectangular domain ${\Omega \subset \R^2}$ and a function ${h : \R^2 \times \R^{\ell} \mapsto \R}$ parameterized by ${\theta_g \in \R^{\ell}}$, we define 
an \textit{implicit topology} as the zero sublevel set of the parametric function $h$, that is: 
\begin{equation}
\label{eq:implicit_topology_definition}
T := \{\mbf{x} | \mbf{x} \in \Omega, \; h(\mbf{x}; \theta_g) \leq 0\}.
\end{equation}
The set $T$ can be interpreted as the subset of the design area in which a target material of non-unit refractive index (e.g., SiO$_2$) is to be placed, where the 
remainder of the area $\Omega \setminus T$ contains some given background/substrate material (e.g., freespace, or a dielectric). 
See the lower right-hand side of \cref{fig:hardware_parameterizations} for an illustration of $h$ (shown as the surface on the far right), and $T$ (shaded in blue).

We discretize the domain, associating the material permittivity/permeability values to each grid cell. 
To determine these values, we use the standard approach in level-set topology optimization: for each grid cell we compute the area of intersection with $T$. 
Then the material properties associated with each grid cell are proportional to the area of intersection. 
While this leads to some intermediate permittivities/permeabilities, these are restricted to the boundary of the topology $\partial T$. 

In the general approach described here, the objective is highly nonconvex and nondifferentiable with respect to the material distribution, so a variety of relaxation approaches are 
employed. 
The usual approach to computing the area of intersection is to smooth the boundary of the topology using a heuristic (e.g., a relaxed Heaviside). 
Instead, we compute the area of intersection using Fiber Monte Carlo (FMC) \cite{richardson2024fiber}. 

In FMC, a collection of line segements or `fibers' are sampled from $\Omega$, and the fibers are used analogously to points in simple Monte Carlo to derive 
an unbiased estimator of the area of intersection. 
This estimator is differentiable even with an exact Heaviside; obviating the need for heuristic smoothing approaches, and resulting in far more stable optimization 
in our experiments. 

Using FMC also enables the use of highly flexible and structured parameterizations of the topology.
For instance, the center design of \cref{fig:hardware_parameterizations} illustrates that we can directly optimize a triangular mesh topology with respect to both the vertices and the 
permittivities associated with each triangle constituting the mesh directly. 
In this context, the parameters $\theta_g$ correspond directly with the vertex positions and triangle permittivities, not the parameters of an implicit topology. 
The outer designs illustrate implicit topologies derived from B-splines (leftmost), and a neural network implicit field (rightmost), in which $\theta_g$ correspond with spline control points, and neural network parameters, respectively. 

Advanced physical layout and fabrication often requires the designer to arrange a collection of strict geometric primitives (e.g., rectangles or other simple polygons), which can be natively supported in our inverse design scheme.

\subsubsection{Differentiable Simulation}
\label{in_silico:diff_sim}

As we describe in detail in \cref{appendix:related_work:photonic_inverse_design}, there exist a variety of special purpose field solvers exploiting the adjoint method dating back decades \cite{georgieva2002feasible}.
There also exist a handful of more general purpose full-wave solvers \cite{Angler,Ceviche} using early research-grade automatic differentiation tools \cite{maclaurin2015autograd}, and several
excellent modern special-purpose tools for specific electromagnetics applications \cite{laporte2019highly}.

We developed a generic differentiable field solver in Jax \cite{jax2018github}, with support for hardware acceleration on GPU, custom sparse numerical linear algebra primitives with correct
sparse vector-Jacobian product implementations, batched simulations (varying the data but holding the hardware/wave operator fixed), and JIT compiled simulation to amortize overhead involved in multiple successive simulations.

\subsubsection{Inverse Design of Optical Classifier}
\label{in_silico:inverse_design}

The aggregate forward computational model with parameters $\theta := (\theta_d, \theta_g)$ of \cref{eq_optical_network_function} is denoted with: 
\begin{equation} 
\hat{f}(\mbf{x}; \theta)
:= D(\mbf{x}; {\theta_d})R(\mathbf{A}(\theta_g)^{-1}\mathbf{b}(\mbf{x})).
\end{equation}
In multiclass classification, we are given a collection of data $\mathcal{D} := \{(\mbf{x}_i, \mbf{c}_i)\}_{i=1}^N$ comprised of inputs $\mbf{x}_i$ with labels $\mbf{c}_i$. 
Then we aim to locally solve the following optimization problem, 
\begin{equation} \label{eq_general_opt_problem}
\text{minimize}_{\theta} \sum_{i=1}^N \mathcal{L}(\mbf{c}_i, \text{Softmax}(\hat{f}(\mbf{x}_i; \theta))),
\end{equation}
with an appropriate objective (e.g., cross-entropy) $\mathcal{L}$.

We use gradient based optimization (specifically, Adam \cite{kingma2014adam}), exploiting FMC-based topology optimization and differentiable simulation to obtain the derivatives with respect to the material distribution parameters $\theta_g$,
and conventional automatic differentiation for the derivatives of the decoder parameters $\theta_d$ with respect to the objective. 

\section{Experiments}
\label{experiments}

Our experimental setup is the inverse design of a two dimensional optical classifier using a regular array of cylindrical (i.e., dipole) sources for our positional encoding (see \cref{fig:system_design}). 
Like prior work \cite{khoram2019nanophotonic,wanjura2024fully}, we consider multi-class classification over the MNIST dataset, which is comprised of $28 \times 28$ grayscale images at $8$-bit precision, each of which belongs to one of 10 distinct classes. 

\subsection{Encoding}

We first linearly project each input \mbox{$\mbf{x} \in \mathbb{R}^{784}$} to a $p$-vector $\tilde{\mbf{x}} \in \mathbb{R}^p$, with values of $p = 4, 8, 16$ evaluated in our experiments. 
The linear projection is derived from principal components analysis (PCA). 
Then, for each of $p$ dimensions comprising $\tilde{\mbf{x}}$, we uniformly quantize the entry into $q$ bins, reducing the precision of the entry to $\log_2q$ bits. 
In our experiments, we use $q = 4, 8, 16$.
These configurations of $(p, q)$ correspond with each image encoded with $8, 24$, and $64$ bits. 
Physically, this encoding corresponds with a $p \times q$ array of identical cylindrical sources laid out within the design area. 
For any given $\tilde{\mbf{x}}$, the $p$\textsuperscript{th} row of the source array has precisely one source activated, corresponding with the bin that index is quantized to. 
\begin{figure}[hbt!]
    \includegraphics[width=0.48\textwidth]{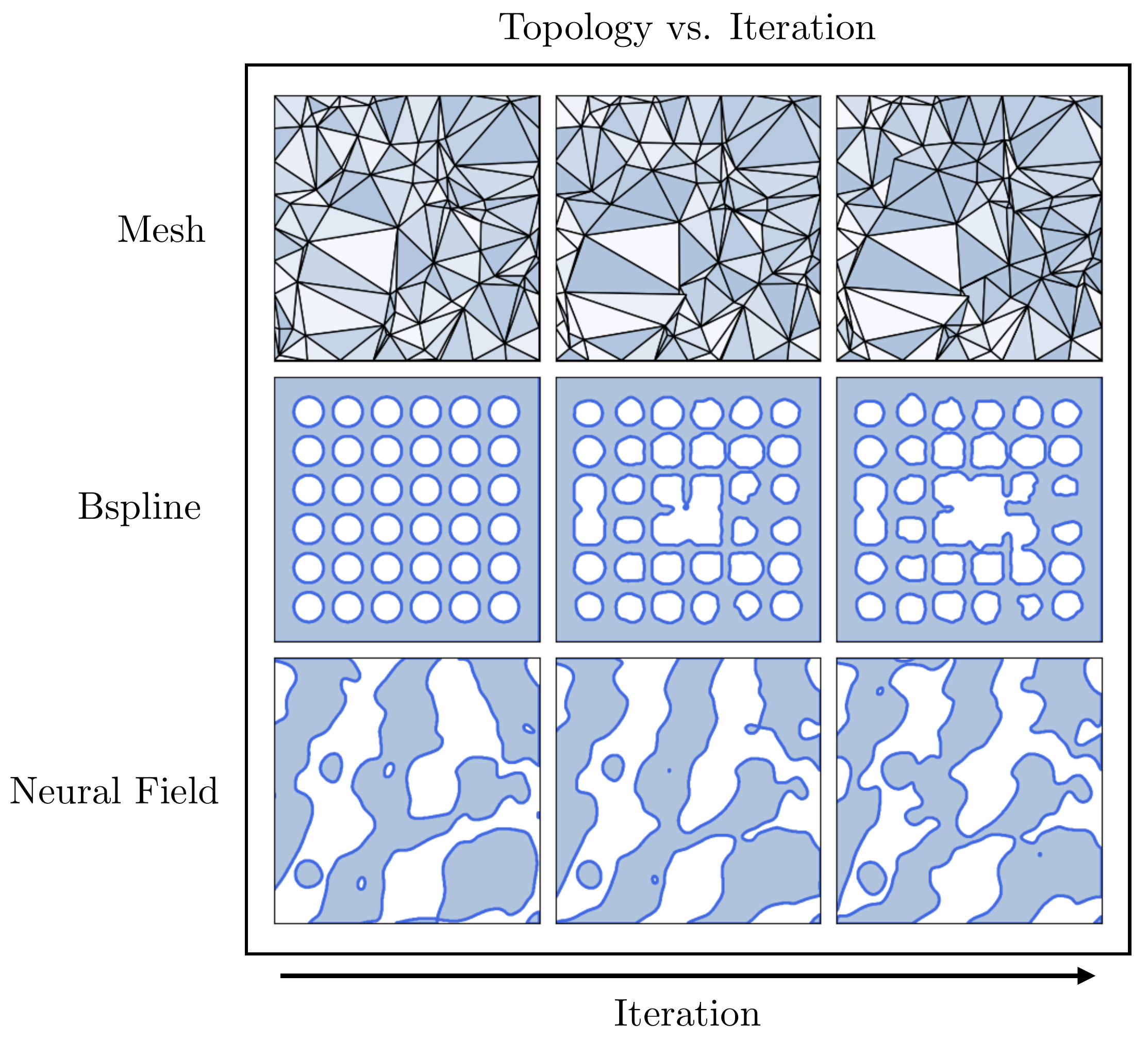}
    \vspace{-0.5cm}%
        \caption{\small Evolution of design topology over the course of optimization. The left-most column shows the initialization of the topology (before optimization), the right-most column shows the final topology (after optimization) and the center column illustrates the topology at an intermediate iteration.}
        \label{fig:topology_evolution}
\end{figure}
\subsection{Ablation}
For each quantization configuration we also fit a purely numerical linear model $\hat{f}(\tilde{\mbf{x}}; \mbf{W}) = \mbf{W} \tilde{\mbf{x}}$ (to the same quantized data), and a standard deep artificial neural network (an MLP), to contextualize the expected performance without using positional encoding; the results are shown in \cref{table:experimental_results}. 

Our optical neural networks significantly improve classification performance compared to a linear model, and achieve performance generally commensurate to the deep artificial neural network. 
The numeric deep network outperforms the optical network more at higher precision. 

We also show examples of how the topology evolves over the course of training/optimization in \cref{fig:topology_evolution}. 
Additional details on experiments and optimization can be found in \cref{appendix:experiments}.

\section{Limitations and Future Work}
\label{limitations_and_future}

\subsection{Limitations}

A limitation of positional encoding is the fact that the nonlinearities hold only elementwise with respect to the data. 
With the simpler tasks often used for evaluation in this space, this does not present a major impediment, but more complicated applications often require nonlinear mixtures of variables. 
Future work should investigate whether this limitation can be overcome by introducing a collection of \textit{layers} of our compute elements, interconnected with nonlinear optical routers. 

A practical bottleneck of optimization in our bottom-up physics based approach is differentiable simulation. 
In the frequency domain, simulation scales (roughly) as $O(n^3)$ with an input domain of size $n \times n$, so scaling to larger designs likely requires reducing calls to the full field solver, possibly using queries to surrogate models or model reduction-based systems \cite{cozad2014learning}. 

\subsection{Future Work}
\cref{eq:nonlinear_field} makes explicit a nonlinear relationship between both the source position and the measured fields, and between the source \textit{frequency} and the measured fields. 
A follow-on work will explore this frequency-based nonlinearity and describe its implications and relationship to the position-based nonlinearity introduced in this work. 

Future physical layer inverse design tools could attempt to directly utilize an FMC-based topology optimization approach to design highly structured geometric layouts like those of a modern digital (or analog) CMOS electronic design, with appropriate modifications to the simulation framework. 

\subsection{Conclusion}

We contribute a novel method for low-power nonlinear computation in fully linear media. 
The proposed source positional encoding is straightforward, requiring only the ability to drive the system at distinct locations. 

We also develop a fully automated inverse design system for extremely specialized optical computing elements. 
Further development of automated hardware design is critical to practically reaping the benefits of hardware specialization moving forward. 
Both positional encoding and automated hardware design generalizes well beyond optics, and may prove to be significant within a variety of computing paradigms. 

\subsection{Acknowledgements}

The authors would like to thank Jake Schaefer and Eric Blow for helpful feedback on the manuscript.

N.R. was partially supported by the National Science Foundation (NSF): OAC-2118201.

C.B. was supported by the Swiss National Science Foundation (SNSF) through a Postdoc.Mobility fellowship (P500PT 217673/1).

\bibliography{cem}
\bibliographystyle{icml2024}

\newpage
\appendix
\onecolumn
\section{Appendix}

\subsection{Notation}

\begin{flalign*}
\nonumber \mathbb{R}_{\succeq 0} \quad \quad \quad &\text{The set of nonnegative real numbers $\{x \in \mathbb{R}\; | \; x \geq 0\}$.}&&\\
\nonumber \mathbb{R}_{\succ 0} \quad \quad \quad &\text{The set of positive real numbers $\{x \in \mathbb{R}\; | \; x > 0\}$.}&&\\
\nonumber \mathbb{T} \subseteq \mathbb{R}_{\succeq 0} \quad \quad \quad &\text{An alias for the nonnegative reals used to suggest a temporal coordinate.}&&\\
\nonumber \partial X \quad \quad \quad &\text{Boundary of the set $X$, i.e., the set difference of its closure against its interior.}&&\\
\end{flalign*}

\subsection{Electromagnetics}
\label{appendix_em}

We follow the notation given in \cite{balanis2012advanced}, which is briefly discussed here and summarized at the conclusion of this section. 
Importantly, we work in the frequency domain, meaning we assume that the fields are harmonic functions of their temporal argument. 
We use boldface font to denote the complex-valued form of a given field/quantity, and script font for the associated time-dependent field/quantity. 
These representations are related through a complex sinuisoidal term $e^{j\omega t}$ where the frequency $\omega$ in units of [radians/s] is implicit (but assumed fixed across all quantities) and $j$ denotes the imaginary unit, as is standard in electrical engineering. 

\begin{flalign*}
\nonumber \mathcal{E} : \mathbb{R}^3 \times \mathbb{T} \mapsto \mathbb{R}^3 \quad \quad \quad &\text{The time-harmonic electric field. If the complex electric field is denoted $\mathbf{E} : \mathbb{R}^3 \mapsto \mathbb{C}^3$,} &&\\
\nonumber &\text{then $\mathcal{E}(\mathbf{x}, t) = \text{Re}(e^{j\omega t}\mathbf{E}(\mathbf{x}))$. Units [V/m].}&&\\
\nonumber \mathcal{H} : \mathbb{R}^3 \times \mathbb{T} \mapsto \mathbb{R}^3 \quad \quad \quad &\text{The time-harmonic magnetic field. If the complex magnetic field is denoted $\mathbf{H} : \mathbb{R}^3 \mapsto \mathbb{C}^3$,} &&\\
\nonumber &\text{then $\mathcal{H}(\mathbf{x}, t) = \text{Re}(e^{j\omega t}\mathbf{H}(\mathbf{x}))$. Units [A/m].}&&\\
\nonumber \mathcal{D} : \mathbb{R}^3 \times \mathbb{T} \mapsto \mathbb{R}^3 \quad \quad \quad &\text{The time-harmonic electric flux density. If the complex electric flux density is denoted} &&\\
\nonumber &\text{$\mathbf{D} : \mathbb{R}^3  \mapsto \mathbb{C}^3$, then $\mathcal{D}(\mathbf{x}, t) = \text{Re}(e^{j\omega t}\mathbf{D}(\mathbf{x}))$. Units [C/$\text{m}^2$].}&&\\
\nonumber \mathcal{B} : \mathbb{R}^3 \times \mathbb{T} \mapsto \mathbb{R}^3 \quad \quad \quad &\text{The time-harmonic magnetic flux density. If the complex magnetic flux density is denoted} &&\\
\nonumber &\text{$\mathbf{B} : \mathbb{R}^3 \mapsto \mathbb{C}^3$, then $\mathcal{B}(\mathbf{x}, t) = \text{Re}(e^{j\omega t}\mathbf{B}(\mathbf{x}))$. Units [W/$\text{m}^2$].}&&\\
\nonumber \mathcal{J} : \mathbb{R}^3 \times \mathbb{T} \mapsto \mathbb{R}^3 \quad \quad \quad &\text{The time-harmonic electric current density. If the complex electric current density is denoted} &&\\
\nonumber &\text{$\mathbf{B} : \mathbb{R}^3 \mapsto \mathbb{C}^3$, then $\mathcal{B}(\mathbf{x}, t) = \text{Re}(e^{j\omega t}\mathbf{B}(\mathbf{x}))$. Units [W/$\text{m}^2$].}&&\\
\nonumber \mathcal{Q} : \mathbb{R}^3 \times \mathbb{T} \mapsto \mathbb{R} \quad \quad \quad \; \; &\text{The time-harmonic electric charge density. If the complex electric charge density is denoted} &&\\
\nonumber &\text{$\mathbf{Q} : \mathbb{R}^3 \mapsto \mathbb{C}$, then $\mathcal{Q}(\mathbf{x}, t) = \text{Re}(e^{j\omega t}\mathbf{Q}(\mathbf{x}))$. Units [C/$\text{m}^3$].}&&
\end{flalign*}

To denote the material parameters which appear in the constitutive relations, we use the following scalar functions of position. 

\begin{flalign*}
\nonumber \epsilon : \mathbb{R}^3 \mapsto \mathbb{C} \quad \quad \quad \; \; &\text{The scalar, complex-valued electric permittivity/conductivity. The real part $\text{Re}(\epsilon)$ is the electric} &&\\
\nonumber &\text{permittivity and the imaginary part $\text{Im}(\epsilon) = \sigma$ is the electric conductivity.}&&\\
\nonumber \mu : \mathbb{R}^3 \mapsto \mathbb{R} \quad \quad \quad \; \; &\text{The scalar, real-valued magnetic permeability.}&&
\end{flalign*}

This choice follows from our material assumptions, which are standard in many computational electromagnetics applications. 
We assume that all materials in our simulations are linear, isotropic, and non-dispersive. 
In electromagnetics a material is \textit{linear} if the constitutive parameters (i.e., the value of $(\epsilon, \mu)$ at any point in the simulation domain) do not depend on the value of the field intensities $(||\mathcal{E}||_2^2, ||\mathcal{H}||_2^2)$. 
A material is \textit{isotropic} if its constitutive parameters do not depend on the directions $(\nicefrac{1}{||\mathcal{E}||} \; \mathcal{E}, \nicefrac{1}{||\mathcal{H}||} \; \mathcal{H})$ of the applied fields. 
Finally, a material is \textit{non-dispersive} if its constitutive parameters do not depend on the frequency $\omega$ of excitation. 

In the general case, the constitutive parameters relate the fields through convolution, but in the frequency domain (and representing the constitutive parameters as scalar values), we can simply write 
\begin{align}
    \D &= \epsilon \E \\ 
    \B &= \mu \H.
\end{align}

We use $\epsilon_0, \mu_0$ to denote the permittivity and permeability of the vacuum, respectively. 
We denote the \textit{relative permittivity} $\epsilon_r = \nicefrac{\epsilon}{\epsilon_0}$ and the \textit{relative permeability} $\mu_r = \nicefrac{\mu}{\mu_0}$. 
We sometimes refer to the relative permittivity as the \textit{dielectric constant}. 
Finally, we use $k = \omega \sqrt{\mu \epsilon}$ to denote the \textit{wavenumber}, which is related to the wavelength via $k = \nicefrac{2 \pi}{\lambda}$ for wavelength $\lambda$. 
In vacuum, we have $k_0 = \nicefrac{\omega}{c_0}$, where $c_0$ is the speed of light in vacuum. 

\subsection{Additional Discussion of Related Work}
\label{appendix:related_work}

\subsubsection{Nonlinear Optics}
\label{appendix:related_work:nonlinear_optics}

Nonlinear optical effects were first predicted in the early 1930s \cite{goppert1930elementarakte} and then observed in the early 1960s at Bell Labs \cite{bayer1970two}. 
Subsequently, a wide body of work has explored nonlinear interactions with respect to intensity, frequency, phase, and polarization \cite{boyd2008nonlinear}. 

Nonlinearity is observed in all media (including the vacuum) above some finite field intensity \cite{Sauter}. 
Unfortunately, it is generally observed only at the extremes of power density, near the breakdown voltage for most materials. 
At a fundamental level, this is a result of the fact that photons (being bosonic particles) do not obey the Pauli exclusion principle. 
This means that in most conditions in free space, the lack of interaction between photons precludes nonlinear optical effects.

By convention, the term nonlinear optics definitionally implies what we call here \textit{materials-derived} nonlinear effects \cite{boyd2008nonlinear}. 

From a physical perspective, the view of nonlinear optics is correct; although in some technical sense one could make non-contradictory statements in the study of 
vacuum optical nonlinearities, practically speaking the science and engineering of optical nonlinearities is that science concerned with the nonlinear response of a 
material to an applied optical field. 
The nonlinearity we pursue in this work (and has been the major focus of the recent work in optical neural networks), is a \textit{logical} nonlinearity associated with implicitly defined mathematical functions 
whose values are determined by the physical states of the system at hand. 
That is, the physical signals with which we associate logical symbols interact linearly, but our logical symbols are related nonlinearly. 

\subsubsection{Nonlinearities in Optical Neural Networks}
\label{appendix:related_work:optical_neural_networks}

With respect to optical neural networks, the pursuit of a mechanism for practical, low-power nonlinearity is a major program.
\cite{xia2024nonlinear} leverage a multiple-scattering cavity to implement a randomized `reservoir' neural network which is used for feature learning on downstream 
tasks like classification and compression.
Notably, is not a trainable neural network, in the sense that the network does not have parameters which can be fit through optimization, but instead can serve as a fixed unsupervised pre-processing/embedding layer for a downstream model/application. 
Our work differs in our explicit treatment of the \textit{optimization} of the optical computing element, via topology optimization of the underlying material distribution. 

\cite{wanjura2024fully} introduce an alternative methodology for nonlinearities. 
The discrete form of Maxwell's equations in the frequency domain can be cast as a 
certain linear system $\mathbf{A} \mathbf{e} = \mathbf{b}$. 
\cite{wanjura2024fully} essentially observe that a nonlinearity can be supported via matrix inversion, i.e., for $x \in \mathbb{R}$, it follows that 
$f(x) := \mathbf{A}(x)^{-1}\mathbf{b}$ can be chosen in a form such that $f$ is nonlinear in $x$. 
This is something like a `structural nonlinearity' in machine learning terminology, and an effective application of the underlying physics to derive a nonlinearity. 
On the other hand, we achieve a nonlinearity in the source term $\mbf{b}$ above, and use the structural parameters in $\mbf{A}$ ahead of time during training to learn effective electromagnetic mode profiles which can be exploited at inference time by $\mbf{b}$. 

\cite{li2024nonlinear} study what could be called the general case of the work presented in \cite{xia2024nonlinear}, providing an analysis of the generic technique of data 
repitition-based strategies. 
The core argument is that the repitition within diffractive volumes (i.e., the optical cavity used in \cite{xia2024nonlinear}) cannot strictly serve as 
optical implementations of most conventional neural network layers, but are still useful in task specific contexts. 
\cite{yildirim2024nonlinear} introduce a technique which is similar to \cite{xia2024nonlinear} but utilizes multiple modulating planes where data is repeatedly embedded. 

\subsubsection{Inverse Design in Photonics}
\label{appendix:related_work:photonic_inverse_design}

Inverse design of electromagnetic devices is widely studied \cite{molesky2018inverse, christiansen2021inverse, georgieva2002feasible}, ranging from fairly application specific microwave circuit 
parameter optimization \cite{georgieva2002feasible} to flexible structural optimization with full-wave field solvers \cite{Angler, Ceviche}. 
We use the adjoint method, which amounts to implicitly differentiating through our full-wave field solver to compute derivatives with respect to the geometry and material properties of our design. 
The adjoint method dates back to the 1960s \cite{pontrjagin1962mathematical}.

In photonics, the use of modern computational methods (e.g., optimization, machine learning and data fitting), has drawn much interest in the design of flexible, complex, and 
unintuitive electromagnetic system designs across a variety of applications \cite{karahan2023deep, so2020deep}. 
These efforts are in the direction of extreme specialization, leveraging the enormity of modern computing resources at our disposal to manage the complexity of low-level physical device design, whereas human designers are at least apparently obligated to employ abstraction and modularity to succeed in the design of complex systems. 

Our automated design methodology also subscribes to the theme of extreme specialization, but our use of topology optimization specifically is most similar to 
\cite{khoram2019nanophotonic}. 
That said, \cite{khoram2019nanophotonic} depend on a material-derived nonlinearity (i.e., the electrical response of a hypothetical optical saturable absorber), and a discrete 
smoothed level set topology optimization method drawn from the computer vision literature \cite{li2010distance}.
Our method obviates material-derived nonlinearities and an exact topology optimization method, which we can differentiate through using Fiber Monte Carlo \cite{richardson2024fiber}. 

\subsection{Additional Experimental Details}
\label{appendix:experiments}

\subsubsection{Does Codesign Matter?}
\label{appendix:experiments_codesign}

During optimization, we track the root mean square (RMS) of the gradient of the objective with respect to (a) the topology parameters and (b) the decoder parameters, in an attempt to understand the relative contributions to the designs. 
In general (but with some exceptions), we find that the using the implicit neural field methods, topology derivatives tend to be between 1 and 3 orders of magnitude larger in RMS than the decoder derivatives; to first order then, the interpretation is that the classification accuracy of the system is around 10-1000x more sensitive to changes in the topology than that of the decoder parameters. 
That said, we have also found that for many initializations of the topology, the system can be trained in something like a `reservoir-mode', updating only the decoder parameters. 
It is challenging to state precisely what properties must obtain for a set of topologies to be amenable to this, but this could be explored in future work. 
At least conceptually, following our modal analysis from \cref{source_positional_encoding}, it is relevant to have a methodology for local optimization of the topology. 

Optimizing without using the decoder at all degrades the system performance significantly. 
Our interpretation of the relatively smaller decoder gradient magnitude follows our hypothesis that the decoder acts as a `permuting' map to account for the arbitrary assignment of class labels to output measurement locations. 
Then, after initialization, optimizing the decoder amounts to merely breaking symmetry in its weights to achieve a better class label to output region assignment. 
From that point, the absolute magnitude of any given decoder weight has minimal effect, since only their relative values are responsible for the label/output region re-mapping. 

\subsubsection{Computing Resources}
\label{appendix:experiments_computing_resources}
All experiments were carried out on a single node Linux server with a 32 CPUs and an Nvidia RTX 3080Ti GPU. 
For efficiency, we precompute all possible source combinations (derived from the finite number of quantized values) and store them in an in-memory cache, and the wave operator 
(less the contribution from the topology) is precomputed and stored as a sparse matrix in memory. 
We execute batched simulations with a batch size of 32, as we determined that batch sizes larger than this only marginally improved performance. 

We use separate optimizers for the topology and decoder parameters, both Adam optimizers with separate step sizes depending on the topology parameterization used (for implicitly formulated B-splines and neural fields, we use $\alpha_g = 0.001$, whereas 
direct mesh parameterization used $\alpha_g = 0.01$).

\end{document}